\documentclass[aip,amsmath,amssymb,reprint,journal=apl,
manuscript=letter,
doi=true,
layout=twocolumn,
letterpaper,
showpacs,
amsmath,amssymb]{revtex4-1}

\usepackage{graphicx}
\usepackage{amsmath}
\usepackage{array,amssymb,bm,dcolumn}
\usepackage{isomath}
\usepackage{soul}
\usepackage{float}
\usepackage[usenames, dvipsnames]{color}

\begin{document}

%\graphicspath{ {Users\chelseydorow\Documents\Butov Group\Perforated Ramp\Figures\ }}
\DeclareGraphicsExtensions{.pdf,.jpeg,.png}

\preprint{AIP/123-QED}

\title{Indirect excitons in a potential energy landscape created by a perforated electrode}

\author{C. J.~Dorow} \email{cdorow@physics.ucsd.edu} \author{Y. Y.~Kuznetsova} \author{J. R.~Leonard} \author{M. K.~Chu} \author{L. V.~Butov} \affiliation{Department of Physics, University of California at San Diego, La Jolla, CA 92093, USA}

\author{J.~Wilkes} \affiliation{School of Physics and Astronomy, Cardiff University, Cardiff CF24 3AA, United Kingdom}

\author{M.~Hanson} \author{A.~C.~Gossard}
\affiliation{Materials Department, University of California at Santa Barbara, Santa Barbara, CA 93106, USA}

\date{\today}

\begin{abstract}
We report on the principle and realization of an excitonic device: a ramp that directs the transport of indirect excitons down a potential energy gradient created by a perforated electrode at constant voltage. The device provides an experimental proof of principle for controlling exciton transport with electrode density gradients. We observed that the exciton transport distance along the ramp increases with increasing exciton density. This effect is explained in terms of disorder screening by repulsive exciton-exciton interactions. \end{abstract}

\maketitle

An indirect exciton (IX) is a bound state of an electron and hole in spatially separated quantum wells (QWs) [Fig.~1(a)]. The spatial separation reduces the overlap of the electron and hole wavefunctions, resulting in IX lifetimes that are orders of magnitude longer than those of direct excitons. Long lifetimes enable the IXs to travel over large distances before recombination~\cite{Hagn95, Butov98, Larionov00, Butov02, Voros05, Ivanov06, Gartner06, Hammack09, Lasic10, Dubin11, Dubin12, Lasic14}. Due to the spatial separation, the IXs also acquire a built-in dipole moment $ed$, where $d$ is the approximate distance between the QW centers. The dipole moment can be explored to control the IX energy: an electric field $F_{z}$ applied perpendicular to the QW plane shifts the IX energy by $E = -edF_z$.~\cite{Miller85} These properties are advantageous for creating excitonic devices and studying the transport of IXs in electrostatic in-plane potential landscapes $E(x,y) = -edF_z(x,y)$. \par

IX transport has been studied in various potential landscapes that were created by laterally modulated voltage $V(x,y)$. These potential landscapes include ramps~\cite{Hagn95, Gartner06, Leonard12}, lattices~\cite{Remeika09, Remeika12}, traps~\cite{Huber98, High09, Schinner11, Kuznetsova15}, circuit devices~\cite{High07, High08, Grosso09, Andreakou14}, narrow channels~\cite{Grosso09, Vogele09, Cohen11}, conveyers~\cite{Winbow11}, and rotating potentials~\cite{Hasling15}. A set of exciton transport phenomena has been observed, including the inner ring in exciton emission patterns~\cite{Butov02, Ivanov06, Hammack09, Dubin12, Stern08, Ivanov10}, transistor effect for excitons~\cite{High07, High08, Grosso09, Andreakou14}, exciton localization-delocalization transition in random potentials~\cite{Butov02, Ivanov06, Hammack09}, lattices~\cite{Remeika09, Remeika12}, and moving potentials~\cite{Winbow11, Hasling15}, coherent exciton transport with suppressed scattering~\cite{High12, Alloing14}, and both spin transport and textures~\cite{High13}. \par
	
\begin{figure}[!hb]
\includegraphics[width=\linewidth, bb = 0 0 365 450]{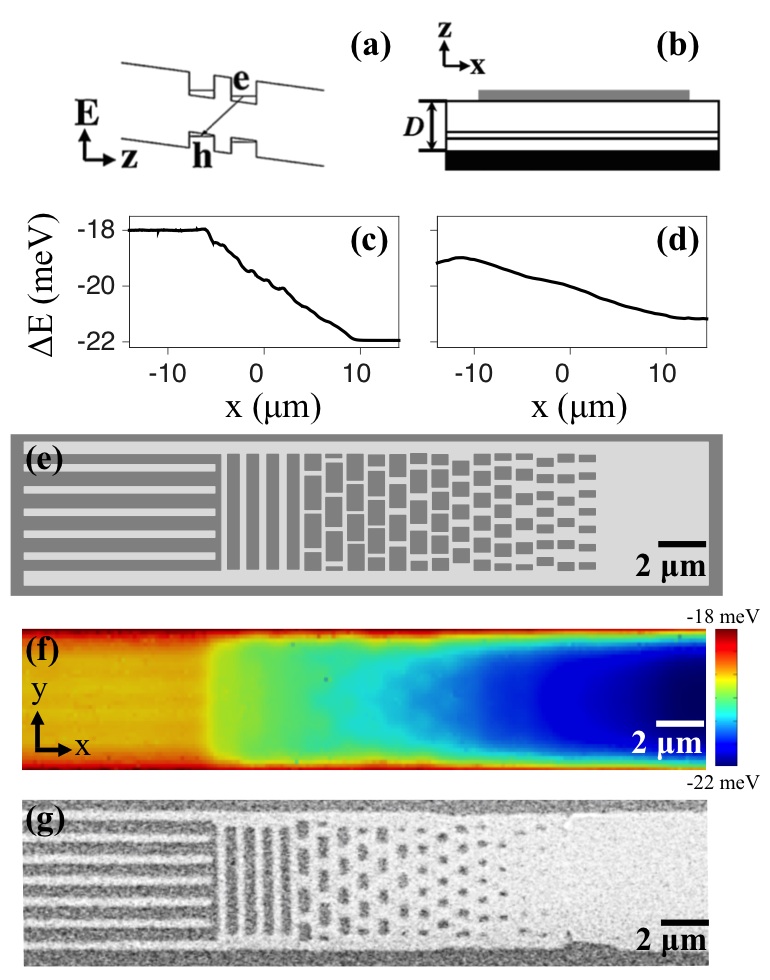} %1500 1800  365 450
\caption{(a) CQW energy band diagram; e, electron; h, hole. (b) Schematic of device. (c) Simulated potential energy $E(x,y) = -edF_z$ for IXs in the ramp. Applied voltage $V = -1.83$~V. (d) IX PL energy in the ramp. $V = -5$ V, $T_{\rm bath} = 1.8$ K, $P_{\rm ex} = 10$~$\mu$W. (e) Ramp electrode (light gray) which forms the potential energy gradient by varying electrode density. (f) Calculated IX energy shift $E(x,y) = -edF_z$ in the ramp. (g) SEM image of the ramp electrode.}
\end{figure}
		
An alternative method for creating potential landscapes for IXs was proposed in Ref.~\cite{Kuznetsova10}, which is based on the lateral modulation of the electrode density rather than voltage. Divergence of the electric field around the periphery of an electrode reduces the normal component $F_z$. This enables one to control $F_z$, and thereby the potential energy landscape for IXs, by adjusting only the electrode density while keeping the entire device at a constant, uniform voltage. This method is beneficial compared to the voltage modulation method because it does not require an energy dissipating voltage gradient. One instance of this method based on varying the electrode width -- the shaped electrode method -- has been used to create confining potentials for IXs in traps~\cite{High09} and ramps~\cite{Leonard12, Andreakou14}. \par
	
In this work, we present an excitonic device based on the electrode density modulation in which a potential energy gradient is created by a perforated electrode at constant voltage. One advantage of the perforated electrode method presented here is the absence of an energy dissipating voltage gradient, similar to a shaped electrode~\cite{Leonard12, Andreakou14}. However, the perforated electrode method also has additional advantages. The shaped electrode method requires a narrow exciton channel limiting the exciton fluxes through the channel. In contrast, the perforated electrode method does not require a narrow exciton channel and allows the width of the channel to be varied independent of the energy gradient. As a result, the perforated electrode method supports larger exciton fluxes than the shaped electrode method. The perforated electrode method gives the opportunity to create versatile potential landscapes for IXs and, in particular, create channels for directing exciton fluxes with the required geometry and energy profile. We present the proof of principle demonstration of the perforated electrode method to control IX energy and fluxes. We realized a potential energy gradient, a ramp, for IXs using a single perforated electrode of constant width and demonstrated IX transport along this ramp. \par

IXs were photoexcited by a 633 nm HeNe laser (focused to a spot with full width half maximum 6~$\mu$m) in a GaAs coupled QW structure (CQW) grown by molecular beam epitaxy [Fig.~1(a,b)]. Exciton photoluminescence (PL) was measured by a spectrometer and a liquid-nitrogen cooled CCD. In the CQW structure, an $n^{+}$-GaAs layer with $n_{\rm Si}$ = $10^{18}$ cm$^{-3}$ serves as a homogeneous ground plane. Two 8 nm GaAs QWs are separated by a 4 nm Al$_{0.33}$Ga$_{0.67}$As barrier and positioned 100~nm above the $n^{+}$-GaAs layer within an undoped $D=1$~$\mu$m thick Al$_{0.33}$Ga$_{0.67}$As layer [Fig.~1(b)]. The CQWs are positioned close to the homogeneous electrode to suppress the in-plane electric field~\cite{Hammack06}, which could otherwise lead to exciton dissociation~\cite{Zimmermann97}. The top semitransparent electrode was fabricated by applying 2~nm Ti, 7~nm Pt, and 2~nm Au.\par
	
Figure~1(e) and 1(g) show a schematic of the electrode pattern and an SEM image of the electrode, respectively. The local electrode density, which controls the normal component $F_z$ and, in turn, the IX energy shift $-edF_z$, is determined by the fraction of the area perforated. The modulation of the electrode density along $x$ is designed to create a ramp potential for IXs with a nearly linear energy gradient [Fig.~1(c),(f)]. Figure~1(c) and (f) show that the perforated electrode creates a smooth ramp potential with the energy variations due to the perforations $\sim 0.1 - 0.3$ meV, significantly smaller than $\sim 1$~meV random potential fluctuations in the CQWs~\cite{Remeika09}. Such energy variations due to the perforations create no substantial obstacles for the exciton transport along the ramp as confirmed by the IX transport measurement described below; perforations create a smooth potential with no significant obstacles for the exciton transport when the lateral dimensions of the perforations are comparable to (or smaller than) the intrinsic layer width $D$ in the device. \par

Figure~1(d) shows the measured spatial profile of the IX energy. It is quantified by the first moment of the IX PL $E(x) = \int EI(x,E)dE/I(x)$, where $I(x) =  \int I(x,E)dE$ is the IX PL intensity. The IX energy also includes the difference between the direct and indirect exciton binding energies, a few meV for the CQWs~\cite{Butov99}. The simulated [Fig.~1(c)] and measured [Fig.~1(d)] spatial profiles of the IX energy are qualitatively similar. However, repulsively interacting IXs screen an external potential which reduces the energy gradient in the ramp [Fig.~1(d)] as discussed below.\par
	
\begin{figure}[h]
\includegraphics[width=\linewidth, bb = 0 0 580 325]{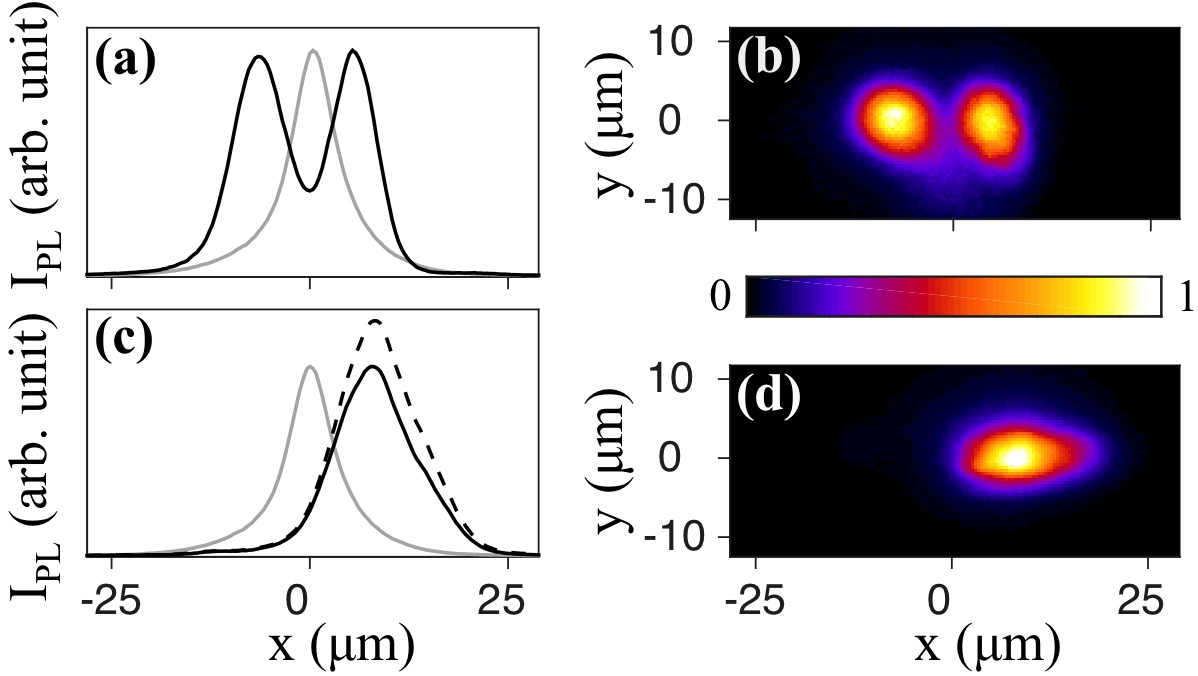}
\caption{(a) IX PL profile (black) in a channel with constant potential energy formed by an electrode of constant density. Laser excitation profile (gray). (b) Image of IX PL intensity in the channel with constant potential energy. In (a) and (b), the channel is created by a 5~$\mu$m wide electrode with no perforations. (c) IX PL profile (solid black) in the ramp. Laser excitation profile (gray). Estimation of IX PL profile (dashed black) with correction for the electrode transparency. (d) Image of IX PL intensity in the ramp. For all data, the laser excitation is centered around $x,y = 0$, $P_{\rm ex} = 10$~$\mu$W, $T_{\rm bath} = 1.8$~K, $V = -5$~V.}
\end{figure}
	
The black curve in Fig.~2(a) shows the profile of the IX PL intensity $I(x)$ in a flat channel with no potential energy gradient. This channel was produced by a solid electrode strip of width 5~$\mu$m with no perforations. The laser excitation profile is shown in gray. Figure~2(b) shows the corresponding spatial distribution of the IX PL. As there is no preferred direction of transport in the flat channel, two maxima that are nearly symmetric relative to the excitation spot position are observed. This phenomenon is consistent with the previously studied inner ring effect \cite{Butov02, Ivanov06, Hammack09, Dubin12, Stern08, Ivanov10}. The inner ring emission is an effect of excitons cooling down to low-energy optically active states \cite{Andreani91} as they travel away from the hot optical excitation spot. Figure~2(c) shows the profile of the IX PL intensity and Fig.~2(d) shows the spatial distribution of IX PL in the ramp. The IX transport in the ramp is directed toward lower IX energy. Such directed transport of IXs in the ramp is analogous to directed transport of electrons in a diode.

Since the electrode is semitransparent and the perforation density varies, the percentage of the signal transmitted through the electrode is not constant along the ramp. The dashed black line in Fig.~2(c) is an estimate of the PL intensity corrected for the varying electrode transparency. This estimation is described in supplementary materials along with other experimental details \cite{supp}. The PL intensity correction due to the electrode transparency variation is quantitative and results in an enhancement of IX transport distance [Fig.~2(c)].~\par
	
\begin{figure}[h]
\includegraphics[width=\linewidth, bb = 0 0 500 405]{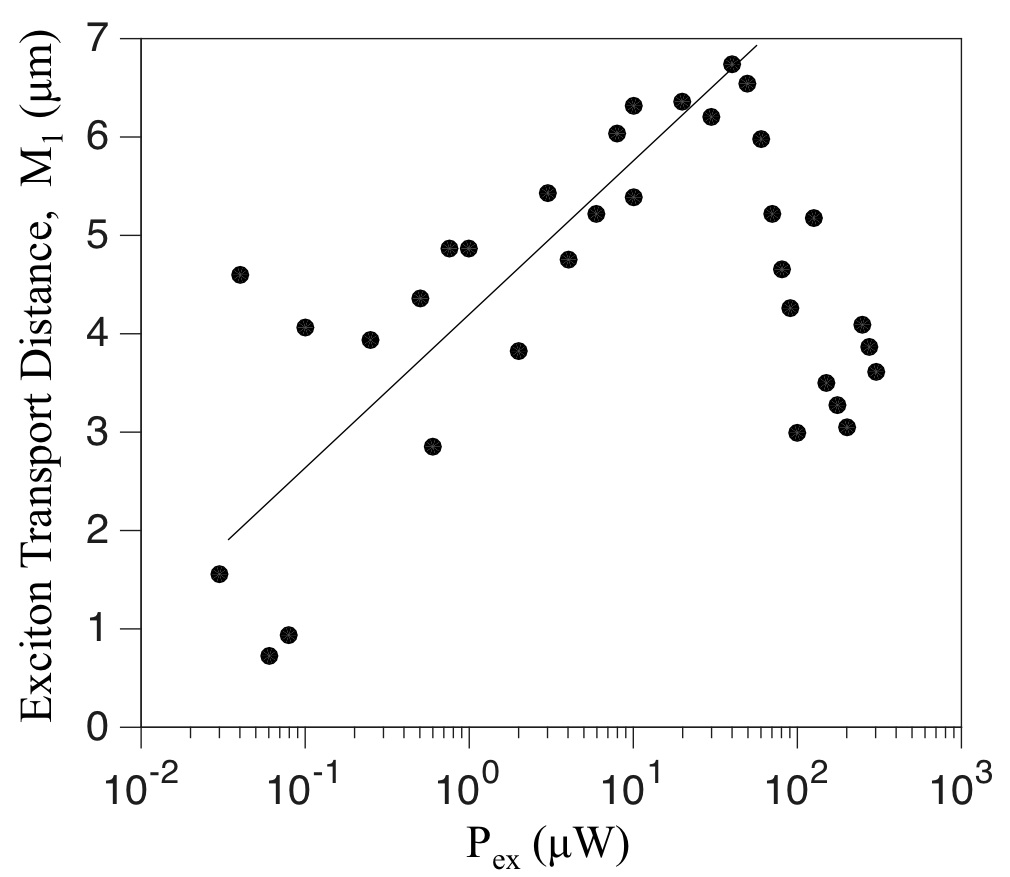}
\caption{The average transport distance of IXs $M_1$ along the ramp as a function of excitation power $P_{\rm ex}$. $T_{\rm bath} = 1.8$~K.}
\end{figure}
	
Figure~3 shows the average IX transport distance, which is quantified by the first moment of the IX PL intensity $M_{1} = \int x I(x)dx/\int I(x)dx$, as a function of laser excitation power $P_{\rm ex}$. The transport distance increases with increasing $P_{\rm ex}$ up to $\sim 50$~$\mu$W, an effect that can be attributed to the screening of disorder in the structure by repulsive exciton-exciton interactions. Repulsive interactions arise from the alignment of the excitons' electric dipole moments perpendicular to the QW plane. Disorder screening improves the IX mobility and thereby increases IX transport distance along the ramp~\cite{Ivanov02, Ivanov06}. This interpretation is supported by a theoretical model presented below. A drop of $M_{1}$ observed at the highest $P_{\rm ex}$ can be related to a photoexcitation induced reduction of $F_z$.
	
\begin{figure}[h]
\includegraphics[width=\linewidth, bb = 0 0 500 585]{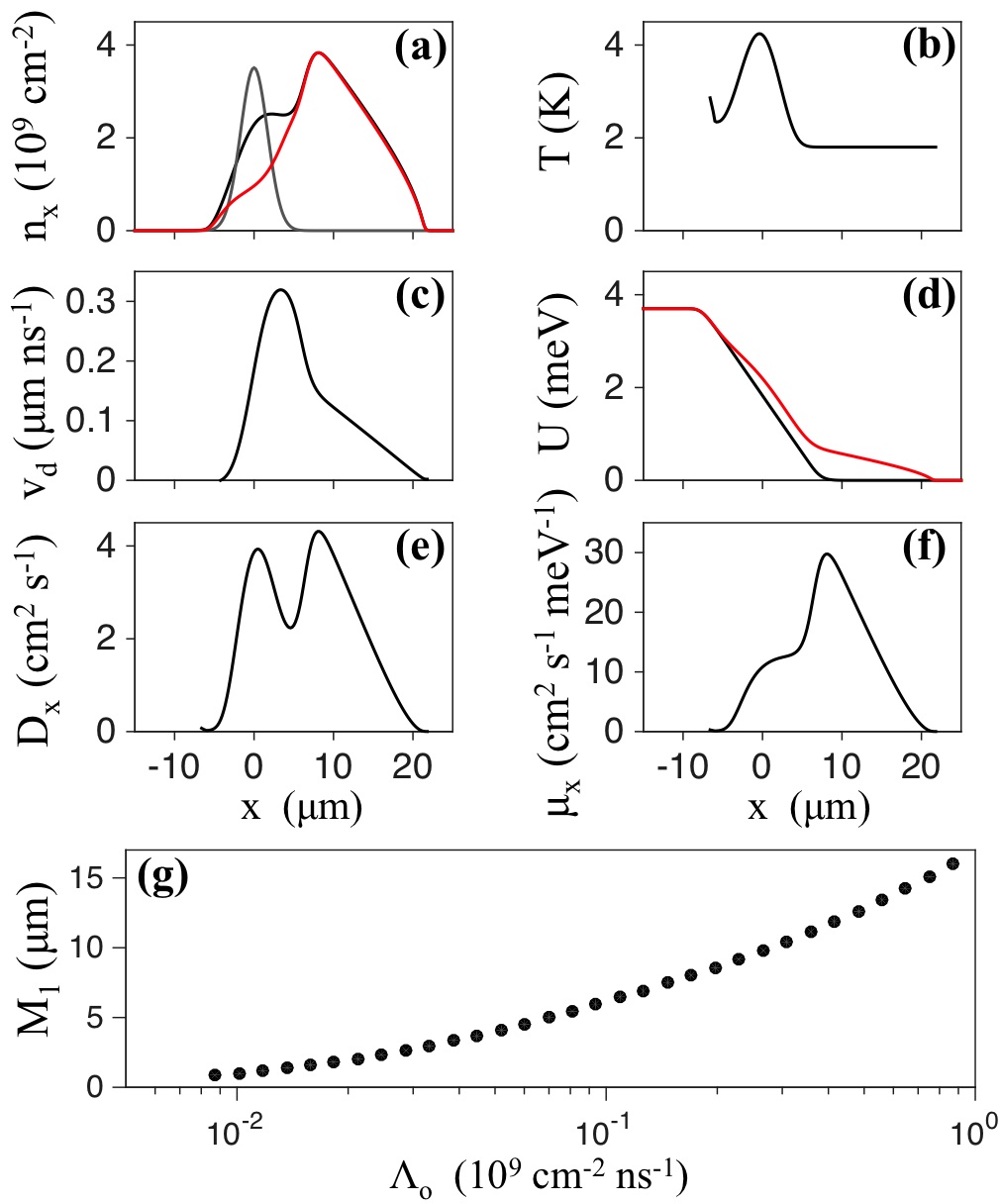}
\caption{Theoretical simulations. (a) IX density (black), IX PL intensity (red), and laser excitation profile (gray). (b) IX temperature. (c) IX drift velocity. (d) Ramp potential (black) and effective IX potential (red) due to the ramp potential plus exciton-exciton interaction. (e) IX diffusion coefficient. (f) IX mobility. $\Lambda_{\rm o} =  0.23 \times 10^9$~cm$^{-2}$ns$^{-1}$ and $T_{\rm bath} = 1.8$~K for simulations (a-f). (g) The average IX transport distance along the ramp as a function of exciton generation rate $\Lambda_{\rm o}$. $T_{\rm bath} = 1.8$~K.}
\end{figure}

The transport of IXs was simulated for a smooth potential energy gradient matching that generated by our device. In this model, the energy variations due to the perforations were neglected since they are small compared with the disorder. Similar to our previous work~\cite{Leonard12, Andreakou14}, the following drift-diffusion equation is solved for the IX density, $n_{\rm x}$:
\begin{equation}
\frac{\partial n_{\rm x}}{\partial t} = \nabla \left[D_{\rm x} \nabla n_{\rm x} + \mu_{\rm x} n_{\rm x} \nabla \left(u_0 n_{\rm x} + U_{\rm ramp}\right)\right] - \frac{n_{\rm x}}{\tau_{\rm opt}} + \Lambda.
\label{driftDiff}
\end{equation}
We treat $n_{\rm x}$ as being homogeneous along $y$ and use $\nabla = \partial/\partial x$, approximately matching the geometry of the ramp. The first and second terms in the square brackets in Eqn.\,(\ref{driftDiff}) are due to IX diffusion and drift, respectively. The diffusion coefficient, $D_{\rm x}$, accounts for the screening of the intrinsic QW disorder potential by repulsively interacting IXs and is given by the thermionic model, $D_{\rm x} = D_{\rm x0} e^{-U_0/(u_0n_{\rm x} + k_{\rm B}T)}$. Here, $U_0/2 = 0.75\,{\rm meV}$ is the amplitude of the disorder potential and $T$ is the exciton temperature. The exciton-exciton interaction potential is approximated by $u_0n_{\rm x}$ with $u_0=4\pi de^2/\varepsilon_b$ where $\varepsilon_b$ is the background dielectric constant and $d = 11.5\,{\rm nm}$ is the static dipole moment, corresponding to the center to center distance of the QWs.~\cite{Ivanov02, Ivanov06, Ivanov10} The drift term in Eqn.\,(\ref{driftDiff}) has contributions from the exciton-exciton interactions and the ramp potential, $U_{\rm ramp} = -edF_z$. A generalized Einstein relationship defines the IX mobility, $\mu_{\rm x} = D_{\rm x}(e^{T_0/T} - 1)/(k_{\rm B}T_0)$ where $T_0 = (\pi\hbar^2n_{\rm x})/(2M_{\rm x}k_{\rm B})$ is the quantum degeneracy temperature. The exciton generation rate, $\Lambda$, has a Gaussian profile whose width matches that of the laser excitation spot. $\tau_{\rm opt}$ is the IX optical lifetime and includes the fact that only low energy excitons inside the light cone may couple to light.~\cite{Andreani91} The IX temperature, $T$, is determined by a thermalization equation,
\begin{equation}
\frac{\partial T}{\partial t} = S_{\rm pump}(T_0,T,\Lambda,E_i) - S_{\rm phonon}(T_0,T).
\label{therm}
\end{equation}
Here, $S_{\rm pump}$ is the heating due to the laser which is characterized by the excess energy of photoexcited excitons, $E_i = 17\,{\rm meV}$. $S_{\rm phonon}$ describes cooling by the emission of bulk longitudinal acoustic phonons. Model parameters and details of the calculation of $S_{\rm pump}$, $S_{\rm phonon}$ and $\tau_{\rm opt}$ can be found in Refs.~\cite{Hammack09, Ivanov02, Ivanov06}. The coupled equations (\ref{driftDiff}-\ref{therm}) describing the IX transport and thermalization kinetics are integrated in time until steady state solutions are obtained. The IX PL intensity is then calculated from the IX decay rate, taking into account the aperture angle of the CCD in the experiment.\par
	
The results of the simulations are presented in Fig.~4. Figure~4(a) shows the IX density and PL intensity profiles. The simulated IX PL intensity profile [Fig.~4(a)] is in qualitative agreement with the experiment [Fig.~2(c)]. Figure~4(b) shows an enhancement of the IX temperature in the excitation spot. Figure~4(d) shows the bare ramp potential $U_{\rm ramp}$ and the ramp potential with screening effects from the exciton-exciton repulsion taken into account $U_{\rm ramp} + u_0n_{\rm x}$. Figures~4(c), 4(e), and 4(f) show the IX drift velocity, diffusion coefficient, and mobility, respectively. Their values are comparable to those for IX transport in devices where no electrode perforation is involved \cite{Ivanov06, Hammack09, Leonard12, Andreakou14}, indicating that the perforations do not cause additional substantial obstacles for IX transport. Figure~4(g) shows the average transport distance of IXs along the ramp $M_1$ as a function of exciton generation rate  $\Lambda_0 = \Lambda(x=0)$. The units of $\Lambda_0$ in Fig.~4(g) can be transformed to the excitation power: $P_{\rm ex}$ $\sim$ $\Lambda_0$$A_{\rm ex}$$E_{\rm ex}$/$\alpha$, where $A_{\rm ex}$ is the excitation spot area, $E_{\rm ex}$ is the energy of a photon in the laser excitation, and $\alpha$ is an assumed IX-excitation probability per photon. We note that for the parameters in the experiment $A_{\rm ex}$ $\sim 30$ $\mu$m$^{2}$, $E_{\rm ex}$ $\sim 2$ eV, and $\alpha$ $\sim 0.01$, $\Lambda_0$ = 10$^{8}$ cm$^{-2}$ns$^{-1}$ transforms to $P_{\rm ex}$ =  1 $\mu$W, and the calculated $M_1$ $\sim 5$ $\mu$m at $\Lambda_0$ = 10$^{8}$ cm$^{-2}$ns$^{-1}$ [Fig.~4(g)] corresponds to the measured $M_1$ $\sim 5$ $\mu$m at $P_{\rm ex}$ =  1 $\mu$W [Fig.~3]. The qualitative similarities between this model and the experimental data [compare Fig.~4(a) with Fig.~2(c) and Fig.~4(g) with Fig.~3] confirms that variations in the ramp potential due to the electrode perforations are small enough to not inhibit the mobility of IXs. We note that the theory gives a nonlinear increase of the IX-transport distance with IX density [Fig.~4(g)], mainly due to the nonlinear dependence of the IX diffusion coefficient on IX density, $D_{\rm x} = D_{\rm x}(n_{\rm x})$, given above. \par

In summary, we report on the realization of an electrostatic ramp for indirect excitons by generating a potential energy gradient with electrode density modulation. The electrode density gradient is achieved by perforating a single electrode of constant width. In contrast to the earlier shaped electrode method, the perforated electrode method does not limit the channel width which, in turn, does not restrict the directed exciton fluxes. The perforated ramp device provides an experimental proof of principle for controlling exciton transport with electrode density gradients. The perforated electrode method gives the opportunity to create versatile potential landscapes for indirect excitons and create channels for directing exciton fluxes with the required geometry and energy profile.

This work was supported by NSF Grant No.1407277. This material is based upon work supported by the National Science Foundation Graduate Research Fellowship Program under Grant No. DGE-1144086. Y.Y.K. was supported by an Intel fellowship. J.W. was supported by EPSRC.

\end{document}

\graphicspath{ {Users\chelseydorow\Documents\Butov Group\Perforated Ramp\Figures\ }}
\DeclareGraphicsExtensions{.pdf,.jpeg,.png}

\preprint{AIP/123-QED}

\title{Supporting Materials for ``Indirect excitons in a potential energy landscape created by a perforated electrode"}
%\thanks{Footnote to title of article.}

\author{C. J.~Dorow} \email{cdorow@physics.ucsd.edu} \author{Y. Y.~Kuznetsova} \author{J. R.~Leonard} \author{M. K.~Chu} \author{L. V.~Butov} \affiliation{Department of Physics, University of California at San Diego, La Jolla, CA 92093, USA}
%-0319

\author{J.~Wilkes} \affiliation{School of Physics and Astronomy, Cardiff University, Cardiff CF24 3AA, United Kingdom}

\author{M.~Hanson} \author{A.~C.~Gossard}
\affiliation{Materials Department, University of California at Santa Barbara, Santa Barbara, CA 93106, USA}

\date{\today}

\maketitle

Figure~S1 shows the average IX transport distance along the ramp, $M_{1}$, as a function of laser excitation power for $T_{\rm bath} = 5.4$ K. Compared to the data measured at $T_{\rm bath} = 1.8$~K in Fig.~3, the IX transport distances at $T_{\rm bath} = 5.4$~K are a few $\mu$m longer for low powers. At low powers and temperatures, the average exciton transport distance is reduced due to the disorder potential trapping a significant fraction of the IXs. At higher temperatures IXs trapped in the disorder potential can be thermally activated and the average IX transport distance increases.

\renewcommand*{\thefigure}{S\arabic{figure}}
\setcounter{figure}{0}
\begin{figure}[h]
\includegraphics[width=3.5in, bb = 0 0 490 410]{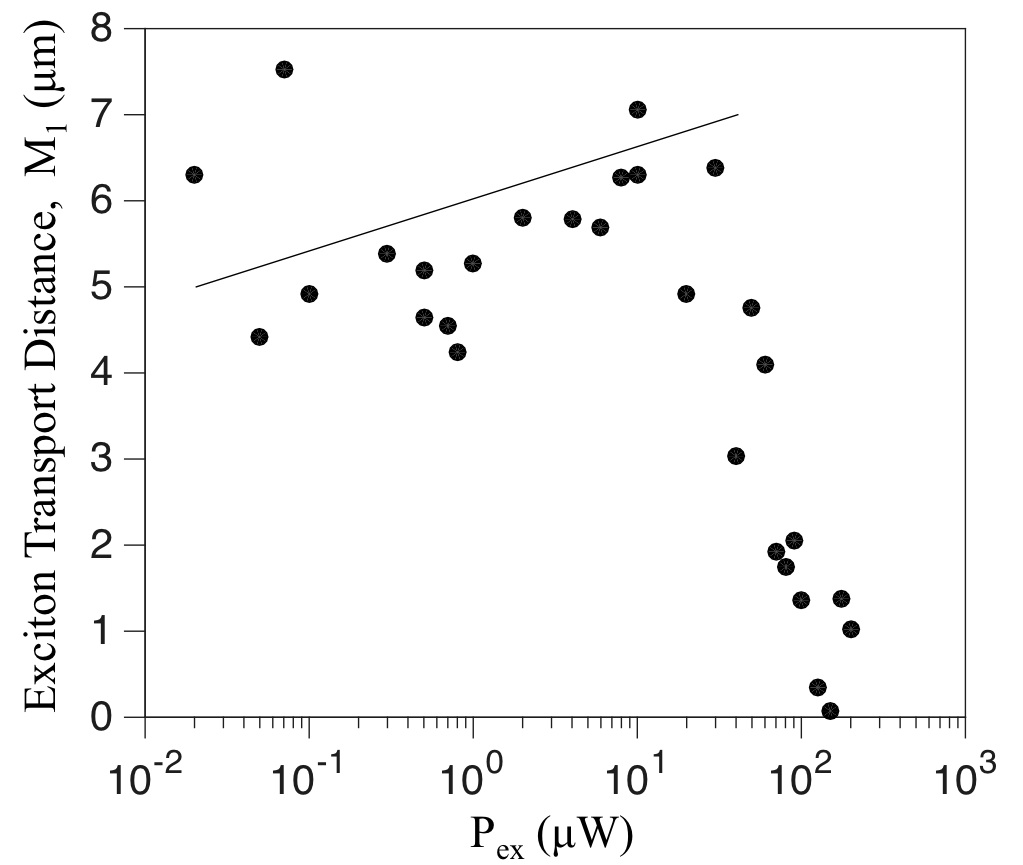}
\caption{Average transport distance of IXs $M_1$ along the ramp as a function of excitation power $P_{\rm ex}$. $T_{\rm bath} = 5.4$~K.}
\end{figure}

Figure~S2 shows an estimation of the average IX transport distance multiplied by a correction factor to account for the electrode transparency. We measured the transparency of the Ti-Pt-Au electrode in the device and found that an electrode without perforations reduces the total PL intensity by about 27~\%. The perforations were taken into account by scaling the correction factor along the ramp by the average local electrode density. The bare measured data in Fig.~3 and corrected data in Fig.~S2 show qualitatively the same trend, but the measured data serves as a lower bound of the efficacy of the ramp as the corrected transport distances are about 1~$\mu$m longer. \par

\begin{figure}[h]
\includegraphics[width=3.5in, bb = 0 0 490 410]{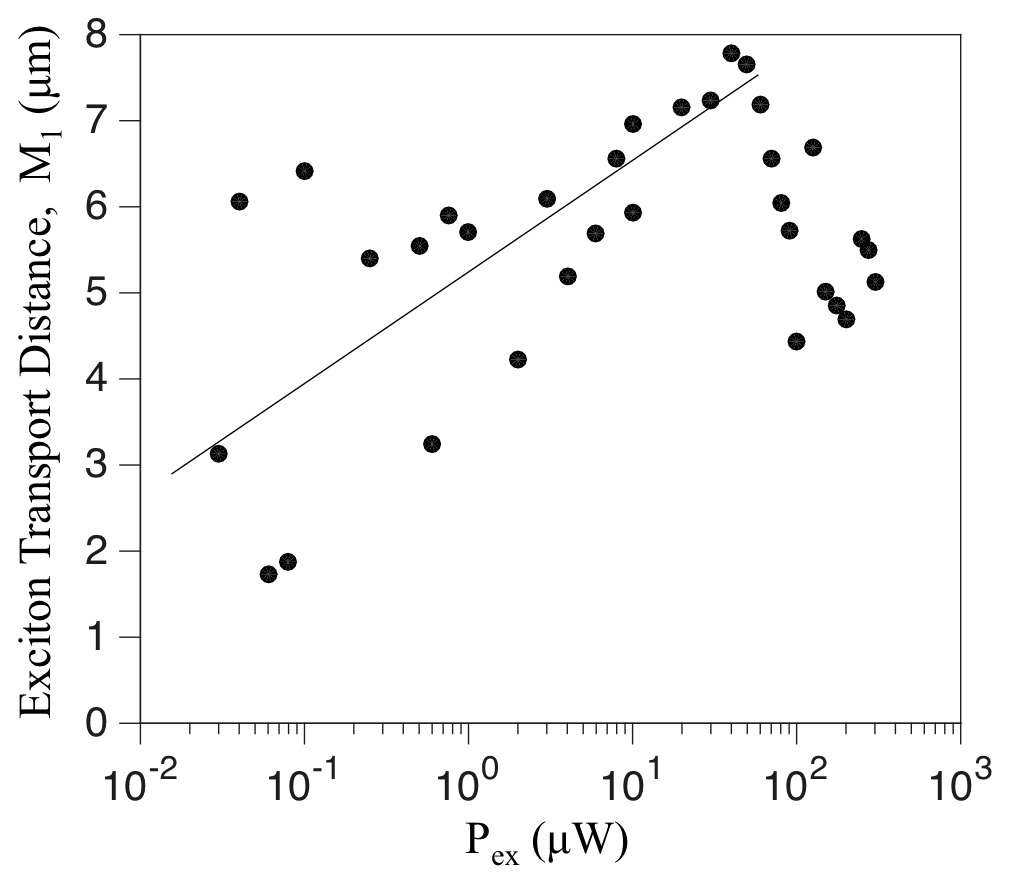}
\caption{Transport distance of IXs $M_{1}$ along the ramp as a function of excitation power $P_{\rm ex}$. $T_{\rm bath} = 1.8$~K. The data are corrected to account for the transparency of the electrode.}
\end{figure}

Figure~S3(a) shows the simulated variations in the potential for transverse cuts across the ramp electrode taken at the positions show in Fig.~S3(b). The largest amplitude of the variations in the potential due to the perforations is $\sim 0.25$~meV (cut b). This is smaller than the estimated amplitude of the disorder potential, 0.75~meV. Therefore, the potential variations due to the perforations are expected to have a negligible effect on IX transport.

\begin{figure}[h]
\includegraphics[width=3.5in, bb = 0 0 420 510]{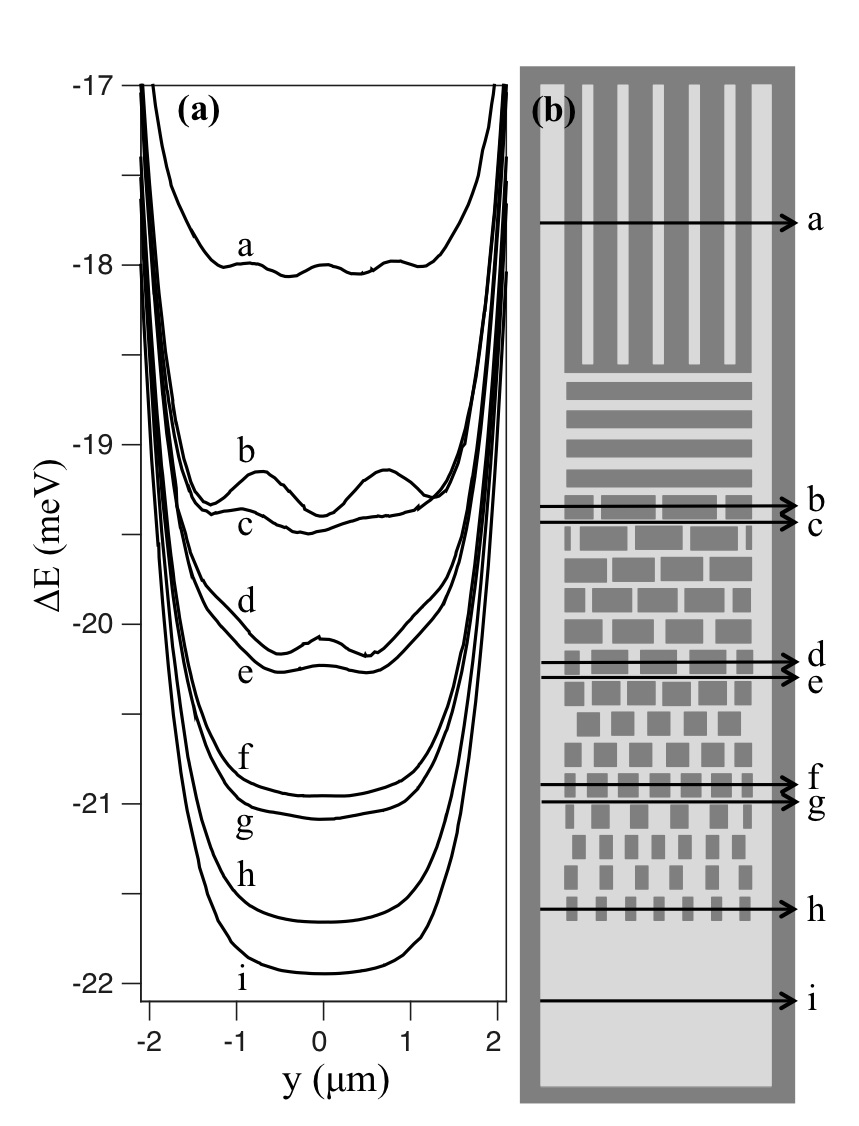}
\caption{(a) Calculated IX potential energy $E(x,y) = -edF_z$ for transverse cuts across the ramp, $V$ = -1.83~V. (b) The positions on the ramp electrode of the transverse cuts.}
\end{figure}

Consistent with the main text, this simulation is done for $V = -1.83$ V which qualitatively agrees with the measured energy shifts for $V = -5$ V. The discrepancy in the simulated and experimental applied voltage is likely due to imperfections in the fabrication of the ramp electrode.